\documentclass{aa}
\usepackage[varg]{txfonts}
\usepackage{graphicx}
\graphicspath{{allsimplots/}}

\usepackage{natbib,twoopt}
\usepackage{lscape}
\usepackage{rotating}
\usepackage[breaklinks=true]{hyperref}
\usepackage{mathtools}
\usepackage[usenames,dvipsnames]{color}
\usepackage[normalem]{ulem} 

\bibpunct{(}{)}{;}{a}{}{,}
\makeatletter
  \newcommandtwoopt{\citeads}[3][][]{\href{http://adsabs.harvard.edu/abs/#3}%
    {\def\hyper@linkstart##1##2{}%
     \let\hyper@linkend\@empty\citealp[#1][#2]{#3}}}
  \newcommandtwoopt{\citepads}[3][][]{\href{http://adsabs.harvard.edu/abs/#3}%
    {\def\hyper@linkstart##1##2{}%
     \let\hyper@linkend\@empty\citep[#1][#2]{#3}}}
  \newcommandtwoopt{\citetads}[3][][]{\href{http://adsabs.harvard.edu/abs/#3}%
    {\def\hyper@linkstart##1##2{}%
     \let\hyper@linkend\@empty\citet[#1][#2]{#3}}}
  \newcommandtwoopt{\citeyearads}[3][][]%
    {\href{http://adsabs.harvard.edu/abs/#3}
    {\def\hyper@linkstart##1##2{}%
     \let\hyper@linkend\@empty\citeyear[#1][#2]{#3}}}
\makeatother

\newcommand{\cmcube}{\,\mathrm{cm^{-3}}}

\newcommand{\Dunitshigh}{\,\times 10^{29}\mathrm{cm}^2\,\mathrm{s}^{-1}}
\newcommand{\Dunits}{\,\times 10^{28}\mathrm{cm}^2\,\mathrm{s}^{-1}}
\newcommand{\Dunitslow}{\,\times 10^{27}\mathrm{cm}^2\,\mathrm{s}^{-1}}

\newcommand{\kpc}{\,\mathrm{kpc}}
\newcommand{\Gyr}{\,\mathrm{Gyr}}
\newcommand{\Myr}{\,\mathrm{Myr}}
\newcommand{\pc}{\,\mathrm{pc}}
\newcommand{\uG}{\,\mu\mathrm{G}}
\newcommand{\Gev}{\,\mathrm{GeV}}
\newcommand{\ev}{\,\mathrm{eV}}
\newcommand{\Mev}{\,\mathrm{MeV}}
\newcommand{\GHz}{\,\mathrm{GHz}}
\newcommand{\MHz}{\,\mathrm{MHz}}
\newcommand{\s}{\,\mathrm{s}}
\newcommand{\yr}{\,\mathrm{yr}}
\newcommand{\creL}{\,l_\mathrm{cr}}

\begin{document}

\title{Modelling the cosmic ray electron propagation in M\,51}

\author{D.\,D.~Mulcahy\inst{1,2}\fnmsep\thanks{david.mulcahy@manchester.ac.uk}
\and A.~Fletcher\inst{3} 
\and R.~Beck\inst{2}
\and D.~Mitra \inst{4,5,6}
\and A.\,M.\,M.~Scaife \inst{1}}

\institute{Jodrell Bank Centre for Astrophysics, Alan Turing Building, School of Physics and Astronomy, The University of Manchester, Oxford Road, Manchester, M13 9PL, U.K
\and Max-Planck-Institut f\"ur Radioastronomie, Auf dem H\"ugel 69, 53121 Bonn, Germany
\and School of Mathematics and Statistics, Newcastle University, Newcastle-upon-Tyne NE1 7RU
\and National Center for Radio Astrophysics, TIFR, Pune University Campus, Ganeshkhind Road, Pune-411007, India
\and Physics Department, University of Vermont, Burlington VT 05405
\and Janusz Gil Institute of Astronomy, University of Zielona G\'ora, Lubuska 2, 65-265 Zielona G\'ora, Poland}

\date{Received 7 March 2016 /
Accepted 2 May 2016}

\abstract{Cosmic ray electrons (CREs) are a crucial part of the ISM and are observed via synchrotron emission. While much modelling has been carried out on the CRE distribution and propagation of the Milky Way, little has been done on normal external star-forming galaxies. Recent spectral data from a new generation of radio telescopes enable us to find more robust estimations of the CRE propagation.}{To model the synchrotron spectral index of M\,51 using the diffusion energy-loss equation and to compare the model results with the observed spectral index determined from recent low-frequency observations with LOFAR.}{We solve the time-dependent diffusion energy-loss equation for CREs in M\,51. This is the first time that this model for CRE propagation has been solved for a realistic distribution of CRE sources, which we derive from the observed star formation rate, in an external galaxy. The radial variation of the synchrotron spectral index and scale-length produced by the model are compared to recent LOFAR and older VLA observational data and also to new observations of M\,51 at 325\,MHz obtained with the GMRT.}{We find that propagation of CREs by diffusion alone is sufficient to reproduce the observed spectral index distribution in M\,51. An isotropic diffusion coefficient with a value of $6.6\pm0.2 \Dunits$ is found to fit best and is similar to what is seen in the Milky Way. We estimate an escape time of $11 \Myr$ from the central galaxy to $88 \Myr$ in the extended disk. It is found that an energy dependence of the diffusion coefficient is not important for CRE energies in the range $0.01\Gev$--$3\Gev$. We are able to reproduce the dependence of the observed synchrotron scale-lengths on frequency, with $l \propto \nu^{-1/4}$ in the outer disk and $l \propto \nu^{-1/8}$ in the inner disk.}{}


\maketitle

\section{Introduction}

Cosmic rays electrons (CREs) are transported through the interstellar medium via two processes, convection in a galactic wind or diffusion. These two process can be distinguished via their CRE propagation length as a function of electron energy \citep{2014A&A...568A..74M}. The propagation length depends upon the ratio of the transport speed, which can differ strongly between diffusive and convective transport, to the CRE lifetime, which is determined by energy loss rates. Diffusion is found to be the most dominant propagation process in the Milky Way \citep{Strong:2011}.

On a macroscopic level, the process of CRE diffusion has been shown to explain the highly isotropic distribution of CREs as well as their retainability within a galaxy. On a microscopic level, diffusion of CREs is the result of their scattering from turbulent magnetohydrodynamic waves and discontinuities of the interstellar plasma. 

Diffusion would also explain why highly charged particles have highly isotropic distributions and why they are well retained in galaxies.
The diffusion of CREs is described by the diffusion energy-loss equation \citep{1994hea2.book.....L,Kardashev1962} and shows the evolution of the energy spectrum as the particles diffuse from their sources. Such losses would include synchrotron and inverse Compton (IC) losses in normal galaxies but Bremsstralung and ionisation losses can be important in starburst galaxies \citep{2010ApJ...717....1L}.

Modelling of the propagation of CREs is important in order to understand the nature of the synchrotron emission observed at different frequencies and to understand if other processes are at work, such as thermal free-free absorption. While thermal free-free absorption is unlikely to flatten the nonthermal spectrum at 333\,MHz \citep{2015MNRAS.449.3879B}, it could start playing a role at even lower frequencies.   With accurate spectral index maps now being produced for external galaxies from a variety of radio telescopes, it is essential that these results are understood with the aid of numerical models.
It is also possible to use these models to help constrain better the magnetic field strength in regions where there is low star formation, such as the extended disk or halo.

\citet{1977A&A....61...59S} studied the radial variation of the spectral index of M\,51 using
analytical solutions of the diffusion loss equation. In his study, several different source functions were used, including a spheroidal source function with unrealistic values for the magnetic field strength of M\,51. In addition to using unrealistic values for the magnetic field strength of M\,51, these kinds of source functions are now dated and a source function extracted from the observed star formation of the galaxy
should be used. Furthermore, advances in computing now enable us to solve the diffusion energy-loss equation via approximation rather than analytical modelling.  
While analytical solutions for simple cases can give insight into the relations between the quantities involved and are good for rough estimates, they can 
become so complicated that no insight is gained \citep{Strong:2011}. On the other hand, numerical models are intuitive because they are able to generate the cosmic ray distribution 
over the galaxy for all species, the best example being the Galactic Propagation (GALPROP) code \citep{1998ApJ...493..694M,1998ApJ...509..212S}.

The value for the diffusion coefficient has been found to be around $D=(4-6) \Dunits$  by fitting diffusion models with direct measurements of CR nuclei (i.e., secondary-to-primary ratios like boron to carbon) within the solar neighbourhood \citep{2001ApJ...547..264J,2002ApJ...565..280M}

While our own Galaxy has been modelled extensively using the GALPROP code, including recent work using the observed synchrotron spectral index to constrain the parameters governing CRE propagation \citep{Strong:2011}, external galaxies have not been modelled with both diffusion and energy losses of CREs since \citet{1977A&A....61...59S}(M\,51), \citet{1978A&A....66..205S}(NGC891) and \citet{1990A&A...239..424P}(NGC4631). Additionally, these were analytical models.
Recently, \citet{2013ApJ...764...37B} theoretically reproduced the perpendicular diffusion coefficient found from radio observations of NGC253 \citep{2009A&A...494..563H}.

\citet{2010ApJ...717....1L} used a steady-state model for the cosmic ray distribution in a homogeneous disk to predict the far-infrared (FIR) radio correlation \citep{1971A&A....15..110V,1973A&A....29..263V}. However, the authors note that in weaker starbursts, where the cooling and escape times are several megayears, the evolution of the CREs via diffusion and convection become important and the steady-state approximation will fail.

Therefore, for a galaxy like M\,51, the diffusion term becomes important and has to be modelled along with the energy losses of the CREs.
Creating a numerical model to solve this equation would enable us to investigate the importance of the different energy loss processes and estimate the magnitude of the
diffusion coefficient for different galaxies. It would also enable us to compare the exponential scale lengths , $I_{\nu} \propto \exp{(-r/l)}$ where $I_{\nu}$ is the synchrotron intensity at frequency $\nu$ and $r$ is the galactocentric radius, to observations and thus indicating if CRE propagation via convection is important in the disk. 
Using the extent of the observed emission as a measure of the size of the synchrotron disk is hampered by the observational sensitivity. Scale lengths at different frequencies provide more useful observational measures of the disk extent to compare to models.

\begin{table}[h!]
\caption{Physical parameters of M\,51=NGC\,5194}
\centering
\begin{tabular}{l l}
\hline\hline
Morphology & SAbc\\
Position of the nucleus & $\alpha(2000)=13^h 29^m 52^s.709$\\
 & $\delta(2000)=+47^\circ 11^\prime 42.59^{\prime \prime}$ \\
Position angle of major axis & $-10^\circ$ ($0^\circ$ is North)\\
Inclination & $-20^\circ$ ($0^\circ$ is face on) \tablefootmark{ a} \\
Distance & 7.6 Mpc \tablefootmark{ b} \\
Optical radius (R$_{25}$) & 3.9$^{\prime}$ (8.6 kpc) \tablefootmark{ c} \\
\hline
\end{tabular}
\tablefoottext{a}{\citet{1974ApJS...27..437T}}
\tablefoottext{b}{\citet{2002ApJ...577...31C}}
\tablefoottext{c}{\citet{2010AJ....140.1194B}}
\label{physicalpara}
\end{table}

In this paper we will model the cosmic ray electron distribution in the disk of M\,51 and thereby compare results to recent observations at numerous frequencies. We will also present a new map of M\,51 observed at 325 MHz with the Giant Meterwave Radio Telescope (GMRT).
In Section~\ref{section2} we will introduce the CRE propagation model that was used to model the diffusion loss equation including inputs such as magnetic field strengths based on observations. In Section~\ref{section3} we shall present the observational data used in this work, the Very Large Array (VLA) maps at 1.4\,GHz and 4.8\,GHz \citet{2011MNRAS.412.2396F}, Low Frequency Array (LOFAR) map at 151\,MHz \citep{2014A&A...568A..74M} and our new GMRT map at 325\,MHz.
Section~\ref{section4} we will investigate the diffusion and injection of CREs in M\,51 with no energy losses. Section~\ref{section5}, we will introduce energy losses to our model and compare the modelled spectral index to the observational spectral index obtained in \citet{2014A&A...568A..74M}. In Section~\ref{section6},  we shall introduce an escape term into our model and determine improved estimations for the diffusion coefficient and escape time. We shall also will investigate the nature of energy dependent diffusion.
Section~\ref{section7} we will compute the observed exponential scale lengths for each observing frequency and compare with our model's results. Finally, Section~\ref{section8} and \ref{section9}, will summarise and discuss the findings of this work as well as predicting the spectral index and  exponential scale lengths of the radio continuum at LOFAR Low Band Antenna (LBA) frequencies.

\section{A cosmic ray propagation model for M\,51  \label{section2}}

We assume that diffusion is the primary means by which the cosmic ray electrons in M\,51 propagate from their sources and that as they propagate they lose energy. The equation describing the distribution of cosmic ray electrons is \citep[Eq. (5.6)]{Berezinskii:1990}
\begin{equation}
\frac{\partial N}{\partial t} = D(E) \nabla^2 N +\frac{\partial}{\partial E}[b(E) N] + Q(\vec{r}, E) - \frac{N}{\tau},
\label{eq:3D}
\end{equation}
where $N(\vec{r}, E, t)$ is the number of particles as a function of position $\vec{r}(r, \phi, z)$ in cylindrical coordinates, energy $E$ and time $t$. The diffusion coefficient $D(E)$ may depend on the energy of the electrons, but is otherwise assumed to be constant. Energy losses, described by the term 
\[
b(E)\equiv -\partial E/\partial t,
\] 
are restricted to synchrotron emission and inverse Compton scattering, the two dominant energy loss mechanisms at the energies of interest since both are proportional to the squared energy of the CRE. Thus we have 
\[
b(E)=\beta E^2
\] 
with $E$ measured in $\Gev$ and \citep[Eq. (5.2)]{Berezinskii:1990}
\begin{equation}
\beta = 8\times 10^{-17}(U_\mathrm{rad} + 6\times 10^{11} B(r)^2 / 8\pi) \Gev^{-1}\s^{-1},
\label{eq:Eloss}
\end{equation}
where $U_\mathrm{rad}\approx 1\ev \cmcube$ is the energy density of the interstellar radiation field and $B(r)$ is the radially averaged magnetic field strength in Gauss. Cosmic ray electrons are injected using the source term $Q(\mathrm{r}, E)$, which may vary with position but is assumed to be constant in time (see Section~\ref{sec:source}). The final term on the right hand side models the escape of electrons from the galaxy on a timescale $\tau$.  

Our aim in this paper is to model the observed radial variation in the spectral index and the  exponential scale length of the synchrotron emission in M\,51. The observational data are averaged in rings, thus removing any azimuthal dependence, and synchrotron intensity is the integrated emission along the line-of-sight, this removing any vertical dependence. So Eq.~(\ref{eq:3D}) can be reduced to a one-dimensional model, with spatial variations only in the radial direction,
\begin{equation}
\frac{\partial N}{\partial t} = \frac{D(E)}{r} \frac{\partial}{\partial r}\left(r \frac{\partial N}{\partial r} \right) +\frac{\partial}{\partial E}[b(E) N] + Q(r, E) - \frac{N}{\tau}.
\label{eq:1D}
\end{equation}
The domain in radius is $0.05\kpc<r<15.05\kpc$ and the cosmic ray electron energies are modelled in the range $0.01\Gev<E<5\Gev$.

We make Eq.~(\ref{eq:1D}) dimensionless by scaling the variables: $t\rightarrow t_0 t'$, $r\rightarrow r_0 r'$, $E\rightarrow E_0 E'$, $B\rightarrow B_0 B'$, $D\rightarrow D_0 D'$ and $Q\rightarrow E^{-\gamma_0}Q_0 Q'$ where primed variables are dimensionless and the subscript $0$ denotes the value used for scaling. This gives
\begin{eqnarray}
\frac{\partial N}{\partial t} & = &
\Theta \frac{D(E)}{r} \frac{\partial}{\partial r}\left(r \frac{\partial N}{\partial r} \right) 
\\ \nonumber
& & + \Phi\frac{\partial}{\partial E}[E^2 N]  
+ \Psi B(r) \frac{\partial}{\partial E}[E^2 N]  
+ Q(r)E^{-\gamma_0} + \frac{N}{\tau},
\label{eq:1Dnodim}
\end{eqnarray}
where all variables are now dimensionless and we have chosen units for the injection such that $Q_0 t_0 E_0^{-\gamma_0}=1$. The dimensionless constants are
\begin{equation}
\Theta = \frac{t_0 D_0}{r_0^2}, \ \ \Phi = 8\times 10^{-17} U_\mathrm{rad} t_0 E_0 , \ \ 
\Psi = 2\times10^{-6} t_0 E_0 B_0^2 . 
\end{equation}
For typical values $t_0=10^9\Gyr$, $r_0=10\kpc$, $E_0=1\Gev$, $B_0=10\uG$ and $D_0=1\Dunits$ we have 
\begin{displaymath}
\Theta=\frac{1}{3}, \ \ \Phi=2.4, \ \ \Psi=6.
\end{displaymath}
Throughout the paper, however, we report all results using dimensional values.

Equation~(\ref{eq:1Dnodim}) was discretised and solved numerically using a 3rd order Runge-Kutta scheme, 4th order spatial derivatives and a 1st order upwind scheme for the energy losses. We imposed a zero-gradient boundary condition on $N(r)$ at the inner boundary and a constant-gradient on $N(r_{max})$ at the outer boundary, where $r_{max}$ = 15.05\,kpc These conditions allow electrons to escape the galaxy at the outer radial boundary, in addition to the losses in the vertical direction modelled by the last term on the right hand side of Eq.~(\ref{eq:1D}). The spatial resolution was $\Delta r = 50\pc$ and we used $300$ logarithmically spaced bins to discretise the cosmic ray electron energy. 
 
The synchrotron intensity $I$ at a given frequency $\nu$ is assumed to be directly proportional to the number of electrons with a particular energy and assumed to emit at their critical frequency, $I_{\nu}(r)\propto N(r, E_{\nu})$, where 
\begin{equation}
E_{\nu} = \sqrt{\frac{\nu_{\mathrm{MHz}}}{16 \,\mathrm{MHz} B_{\perp}}}\mathrm{GeV},
\label{eq:nuEnergy}
\end{equation}
with $B_{\perp}$ the magnetic field strength in the plane of the sky in $\uG$, taken from the magnetic field strength profile shown in Fig.~\ref{bfieldsnrplot} (left). Since we only consider relative intensities derived from the model --- the spectral index between two frequencies and the variation of emissivity with radius --- we do not need to calculate the synchrotron intensity in physical units.
 
\subsection{The distribution of cosmic ray electron sources in M\,51}
\label{sec:source}

A CRE source distribution, $Q(r,E)$, is required by Eq.~(\ref{eq:1D}). Since supernova remnants (SNR) are the most likely origin of the CREs and number of SNR is governed by the star formation rate, we shall use the radial distribution of the star formation rate of M\,51 to define $Q(r)$. In the Milky Way, \citet{2007ARNPS..57..285S} showed that cosmic ray sources are located near the Galactic disk and have a radial distribution similar to that of the SNR.
 
The radial distribution of the star formation rate for M\,51 of  \citet{2007ApJ...671..333K}  was used to infer the radial profile of the supernova rate. \citet{2007ApJ...671..333K} used a combination of H$\alpha$ and 24\,$\mu$m data, which gives reliable extinction corrected ionising fluxes and hence the recent (say within the last $10\Myr$) star formation rate. This decreases rapidly from the central region out to $2\kpc$ with the interarm regions seen as dips at radii of $1$, $2$ and $4\kpc$. The data of \citet{2007ApJ...671..333K} only extends to $r=8\kpc$. For $r>8\kpc$ we make a linear extrapolation from the star formation rate at $r=7.5\kpc$ to zero. Since the star formation rate is already extremely low at $r=7.5\kpc$ this means that effectively our model will contain no CRE sources for $r>8\kpc$.

To determine the supernovae (SNe) rate, we use the multiple part power-law initial-mass-function \citep{2001MNRAS.322..231K} using masses from 0.08  M$_{\odot}$ to 120  M$_{\odot}$. We set our maximum mass to 120  M$_{\odot}$
as this is the upper limit found  where there exists the possibility of the destruction of the star from thermal pulsations \citep{1959ApJ...129..637S,1994ApJS...95..517B}.

We assume that stars with $M> 8 M_{\odot}$ become SNes and so the fraction of mass that goes into SNes is thus 25$\%$.

\begin{figure}
\begin{center}
\includegraphics[scale=0.45]{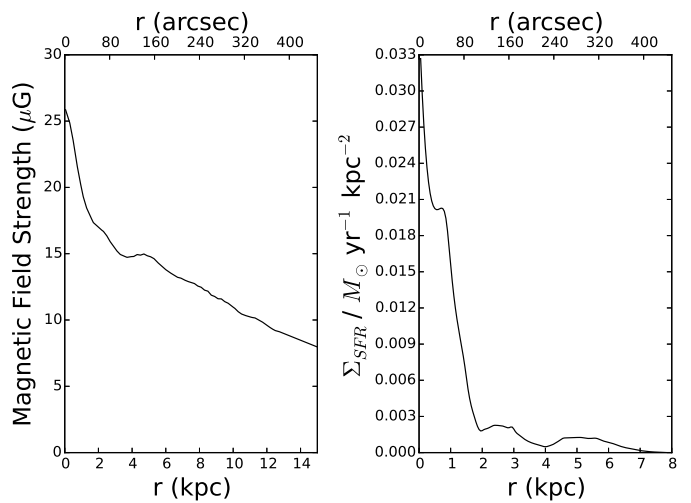}
\end{center}
\caption{Radial profile of the magnetic field strength (left plot) and star formation rate (right plot). These profiles are used in our model to determine the synchrotron losses, via $B(r)$ in Eq.~(\ref{eq:Eloss}) and source distribution for the CREs, $Q(r)$ in Eq.~(\ref{eq:1D}).}
\label{bfieldsnrplot}
\end{figure}

This fraction is applied to the star formation rate (SFR) radial profile of \citet{2007ApJ...671..333K} (Fig.~\ref{bfieldsnrplot}) to find the radial distribution of SNe mass per year.

This model predicts that $1.2 M_{\odot}\yr^{-1}$ are converted to SNe in M\,51. Taking the mean mass of a SNe to be $12 M_{\odot}$, M\,51 should produce on average 1 SNe per 10 years. For comparison the Milky Way produces about 1 SNe per 50 years and one expects M\,51 to have a greater SNe rate due to its higher star formation rate. Furthermore it should be noted that 3 SNe have been detected in M\,51 in the last 20 years (1994I \citep{1994IAUC.5961....1P}, 2005cs \citep{2005MNRAS.364L..33M}, 2011dh \citep{2011ApJ...742L..18A}). In this model, this injection profile is normalised. 

Additionally, we assume that the primary CREs are injected into the galaxy with a power-law spectra $Q(E) = CE^{-\gamma_0}$ with the initial slope of the energy spectrum $\gamma_0=2.0$. This corresponds to $I(\nu)=D\nu^{-0.5}$, where $I(\nu)$ is the synchrotron intensity at frequency $\nu$ which is observed for shell type supernova remnants in the Milky Way \citep{2001AIPC..558...59G}.  $\gamma_0=2.0$ is the standard value for the theory of diffusive shock acceleration in the case of high Mach numbers \citep{bell1978a,bell1978b}. In our model, CREs will be injected at all times and at all radii but weighted with SFR(r) as shown in Fig.~\ref{bfieldsnrplot} (right). 2\% of the SNe energy is assumed to be accelerate CREs.

\subsection{The distribution of the magnetic field strength
\label{sec:bfieldprofile}}

The profile of total magnetic field strength, needed for modelling the synchrotron losses was determined from the total intensity emission of the LOFAR 151\,MHz image \citep{2014A&A...568A..74M} by assuming equipartition between the energy densities of cosmic rays and magnetic field, using the revised formula of \cite{2005AN....326..414B}. The total magnetic field strength scales with the synchrotron intensity I$_{syn}$ as:
\begin{equation}
B_{tot,\perp} \propto I_{syn}^{\,\,\,1/(3-\alpha)},
\end{equation}
where B$_{tot,\perp}$ is the strength of the total field perpendicular to the line of sight. Further assumptions were required such as the synchrotron spectral index of $\alpha$ = -0.8 and the effective pathlength through the source of $1000\pc /\cos{i}=1060\pc$, where $i$ is the inclination of the galaxy. It was also assumed that the polarised emission emerges from ordered fields with all possible inclinations. Here we assume a ratio of CR proton to electron number densities of $K_0 = 100$, which is a reasonable assumption in the star-forming regions in the disk \citep{1978MNRAS.182..443B}. Large uncertainties for the pathlength and $K_0$ of a factor of 2 would effect the result only by 20\%. The effect of adjusting $\alpha$ to between -0.7 and -0.9 produces an error of 5\% in magnetic field strength. Using these assumptions, we created an image of the total magnetic field strength. The AIPS task 'IRING' was used to to create a radial profile of magnetic field strength, taking the inclination and position angle of the galaxy into account.The radial profile of our magnetic field strength in $\mu$G, is shown in Fig. ~\ref{bfieldsnrplot}(left).

\section{Observational data  
\label{section3}}

In our analysis we will use observational data of M\,51 from several different frequencies.
These include VLA observations covered in \citet{2011MNRAS.412.2396F} at 1.4\,GHz \& 4.8\,GHz and LOFAR High Band Antenna (HBA) observations (151\,MHz) from \citet{2014A&A...568A..74M}. Details of these observations and their data reduction are described in the respective papers. 
In addition to these observations we shall use GMRT data at 325\,MHz. Details of the observation and subsequent data reduction are outlined below.

\subsection{GMRT 325\,MHz observation and data reduction}

M\,51 was observed in full synthesis mode with the Giant Meterwave Radio Telescope (GMRT) at 325 MHz. 
GMRT is a full aperture synthesis telescope located near  Pune, India \citep{1990IJRSP..19..493S}. 
The GMRT consists of 30 steerable 45 m diameter dishes over an area of 25 km, and operated at six 
frequencies 150, 230, 325, 610 and 1420 MHz.
The primary beam of the GMRT at 325\,MHz of our observation 
is about 810.$\arcmin$, which is approximately three times larger than 
the 14$\arcmin$ size of M\,51. This enables us to map M\,51 with one pointing at 325 MHz. 
Furthermore, GMRT has 14 antennas located in a central 1 square km, which provides
dense UV coverage for short spacing. On the other hand the largest baseline 
of GMRT is about 25 km, thus enabling high angular resolution.  
The 325 MHz observations were performed on 25/11/2006 using the old hardware correlator backend 
at the GMRT. The 32 MHz bandwidth was split into two sidebands of 16 MHz 
each. We used the higher frequency sideband centred at 333 MHz.

The raw visibilities in Flexible Image Transport Files (FITS) format were used to analyse these data. 
The standard routines of  AIPS (Astronomical Image Processing System)\footnote{http://www.aips.nrao.edu/} were used to reduce these data. The calibrator 3C286 was used 
both as flux and phase calibrator. The flux scale of Reynolds (1994)\footnote{http://www.atnf.csiro.au/observers/memos/d96783~1.pdf} was used. In the full synthesis 
run of approximately 8 hours, 2 hours were spent on the calibrator and 
6 hours on M\,51. The calibrator data
were flagged for both bad baselines and radio frequency interference and were subsequently
used to get the band pass solution and time based gain solutions for each antenna.
These solutions were then applied to the source, M\,51. Data on source
were further edited for interference and the iterative procedure of imaging 
(using facets) and phase only self calibration was performed several times until a good 
point source model of the field was established. At the last stage both amplitude and
phase calibration were applied to the dataset to get the final maps. After flagging, the final image was made 
with 12 MHz bandwidth.

We first cleaned the point sources from each facets in the image using the BGC clean algorithm \citep{1980A&A....89..377C}
and finally the diffuse emission of M\,51 was cleaned using the SDI algorithm \citep{1984A&A...137..159S}.
A Briggs weighting with robust parameter of 0 in 
AIPS was used. Finally the primary beam correction was applied.

\begin{figure}[h!]
\begin{center}
\includegraphics[scale=0.45]{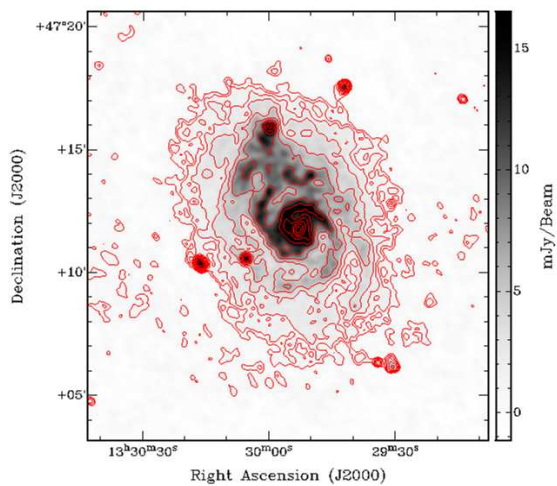}
\end{center}
\caption{M\,51 at 325 MHz observed with the GMRT. Contours are from 900 $\mu$Jy/beam to 90 mJy/beam in steps of 1.5 $\times$ 900 $\mu$Jy/beam. }
\label{GMRTimage}
\end{figure}

The final image is shown in Fig.~\ref{GMRTimage} at 15$\arcsec$ resolution, having an rms noise of 
200 $\mu$Jy/beam. The disk extends out further than seen at 1.4\,GHz but not as far as seen in 151\,MHz \citep{2014A&A...568A..74M} which is expected due to synchrotron losses being greater at higher frequencies. Like the 151\,MHz map, the disk does not extend out uniformly in radius with an excess of emission seen to the north of the galaxy most likely due to the interaction with NGC\,5195. We also see a hint of the HI tidal tale seen by \citep{1990AJ....100..387R}. However, deeper observations would be needed to confirm this. 
The arm and interarm regions are easily seen at 325\,MHz which is in stark contrast to NGC\,6946 at the same frequency \citep{2012MNRAS.419.1136B}. M\,51, which has stronger magnetic fields than most galaxies
due to its interaction with NGC\,5195 causes the CREs to lose their energies on a shorter timescale than in NGC\,6946.

\section{Modelling the radial diffusion of CREs only \label{section4}}

We expect that all of the terms in Eq.~(\ref{eq:1D}) --- diffusion, energy losses, source properties and the escape term --- will be needed to obtain a good model for the observed radial distribution of the synchrotron spectral index. However, we will build up the complexity of the model gradually, in order to develop a better understanding of the role that the different physical properties and processes play. Until energy losses are implemented, in Sect.~\ref{section5}, we obviously will not see any spectral index changes.

\subsection{Diffusion with a single injection of cosmic ray electrons}

First, let us consider the case of single injection of CREs, which then diffuse away from their sources.

\begin{figure}[h!]
\begin{center}
\includegraphics[scale=0.45]{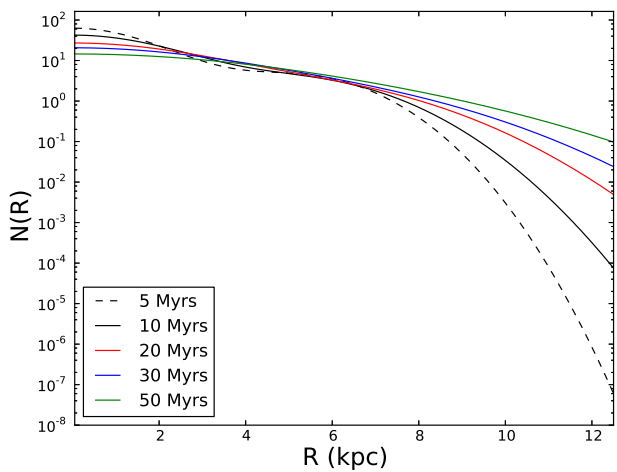}
\end{center}
\caption{The change in the distribution of CREs at energies corresponding to an emission frequency of 1.4\,GHz with time for a diffusion only model. A single injection of CREs, $Q(r)$, occurred at $t=0$. $N(R)$ is in arbitrary units.}
\label{diffusionsingleinjection}
\end{figure}

Fig.~\ref{diffusionsingleinjection} shows a single injection of CREs with no energy losses, using the source term $Q(r)$  based on the $SFR(r)$ shown in Fig.~\ref{bfieldsnrplot} at $t=0$ with the diffusion coefficient set to be  $D_r = 2.8 \Dunits$, which is a similar value to the Milky Way  \citep{2007ARNPS..57..285S}. As expected the CRE distribution throughout M\,51 becomes increasingly uniform.  We see the initial difference between the arm ($r=5.5\kpc$)  and interarm ($r=4\kpc$) CRE density become smeared out by diffusion within $10\Myr$.  This is naturally reflected in the increase of the CRE exponential scale length $\creL$, $N\propto \exp{(-r/\creL)}$, with the scale length increasing from $1.08\kpc$ at $10\Myr$ to $3.32\kpc$ at $80\Myr$. A scale length of $\creL\approx 3\kpc$ is compatible with the observed values for M\,51, which are discussed in more detail in Sect.~\ref{sec:ScaleLength}.
The steep decrease in $N(r)$, which begins at approximately $7\kpc$, is due to the absence of sources in the $SFR(r)$ (Fig.~\ref{bfieldsnrplot}) for $r>7\kpc$. This gradually disappears with time, as the CREs diffuse to larger radii.

\begin{figure}[h!]
\begin{center}
\includegraphics[scale=0.45]{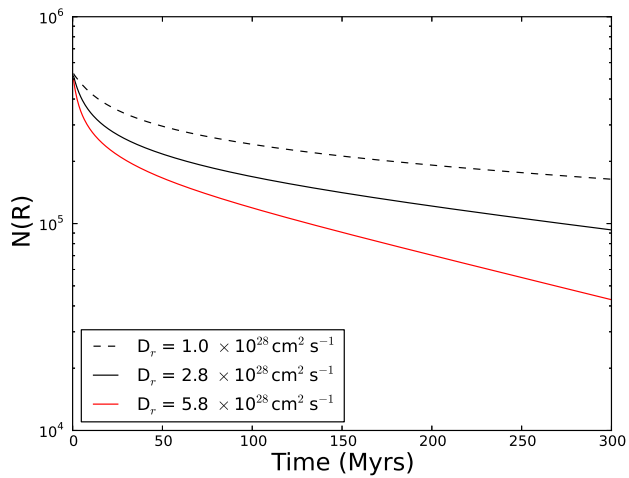}
\end{center}
\caption{The decrease in the total number of CREs with time, for a single injection of CREs at t=0. $N(R)$ is in arbitrary units.}
\label{CREvolutionD}
\end{figure}

The CREs eventually leave the galaxy at $r=r_{max}$ resulting in a decrease in the total number of CREs with time as shown in Fig.~\ref{CREvolutionD}.


\begin{figure}[h!]
\begin{center}
\includegraphics[scale=0.45]{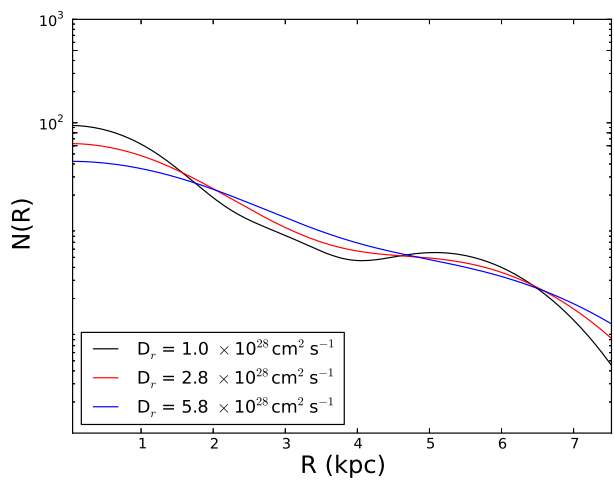}
\end{center}
\caption{The distribution of CREs at energies corresponding to an emission frequency of 1.4\,GHz after 5 Myr while varying the diffusion coefficient with a single injection of CREs. $N(R)$ is in arbitrary units.}
\label{diffusiondiffD}
\end{figure}

The effect of varying the diffusion coefficient on the CRE radial distribution is easily seen in Fig.~\ref{diffusiondiffD}. For a greater value of $D_{r}$, the interarm and arm contrast of M\,51 disappear more quickly with time. When $D_{r} = 5.8 \Dunits$, after $5\Myr$, the interarm and arm contrast of the initial injection of CREs has nearly disappeared. On the other hand, with $D_{r} = 1.0 \Dunits$, the arm, interarm contrast is still apparent.

Another effect of increasing the diffusion coefficient is the speed in which the CREs are able to diffuse out of the galaxy itself. Fig.~\ref{CREvolutionD} shows the decrease of the total number of CREs at all energies of our model for $300\Myr$ after a single injection was implemented at t=0. 
Naturally our model shows the fastest decrease of CREs for greater diffusion coefficients. The total number of CREs would be expected to decrease to zero but the rate of decrease will lessen as the gradient of the radial distribution stabilises.


\subsection{Diffusion with continuous injection of cosmic ray electrons}

One would expect the star formation to change in M\,51 as its companion galaxy, NGC\,5195, has likely passed through the disk of M\,51 several times \citep{2003Ap&SS.284..495T, 2000MNRAS.319..377S}. Several numerical investigations have shown that interactions and mergers can trigger a strong nuclear starburst \citep{2006MNRAS.373.1013C,2000MNRAS.312..859S}. It is worth noting that this may not be always the case as the strong tidal interactions can remove large quantities of gas from the galaxy disks ejected to tidal tails as seen in HI for M\,51. In addition to this, \cite{2007A&A...468...61D} showed that on average, galaxies that pass too close to one another produce the lowest bursts of star formation. Therefore a full time dependent injection function for CREs requires a model for  the SFR history, which is out of the scope of this paper. Instead, we consider a continuous, time-independent injection of CREs. This continuous injection is defined as a single injection but just reapplied at each timestep.

\begin{figure}[h!]
\begin{center}
\includegraphics[scale=0.45]{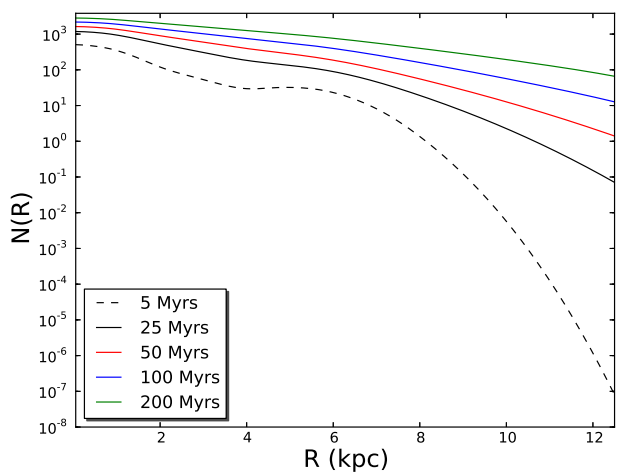}
\end{center}
\caption{The change of the radial distribution of CREs at energies corresponding to an emission frequency of 1.4\,GHz in time with diffusion and a continuous injection of CREs. $N(R)$ is in arbitrary units.}
\label{diffusioncontinuousinjectiondifferenttime}
\end{figure}

Fig.~\ref{diffusioncontinuousinjectiondifferenttime} shows the increasing number of CREs in the galaxy with time. In the extended disk, $r>7.5\kpc$, where the source injection is zero, the change is most dramatic as the inner disk CREs diffuse into this region.
Fig.~\ref{diffusioncontinuousinjectiondifferenttime} shows that the CRE population continues to increase with time since fewer CREs leave the galaxy at the outer boundary, $r_{max}=15.05\kpc$, than are injected at each timestep via $Q(r)$, and so an equilibrium is not reached. The electrons will need to leave the galaxy via a second process in order for this to take place and this will be addressed later in this paper.

With continuous injection, the CRE exponential scale length is $\creL=0.92\kpc$ at $10\Myr$ and $\creL=2.19\kpc$ at $80\Myr$. This is shorter than in the case of a single injection of CREs, since the addition of new CREs at each timestep emphasises the influence of the source distribution and reduces the influence of diffusion. These scale lengths are significantly shorter than the observed values (Sect.~\ref{sec:ScaleLength}).

\begin{figure}[h!]
\begin{center}
\includegraphics[scale=0.45]{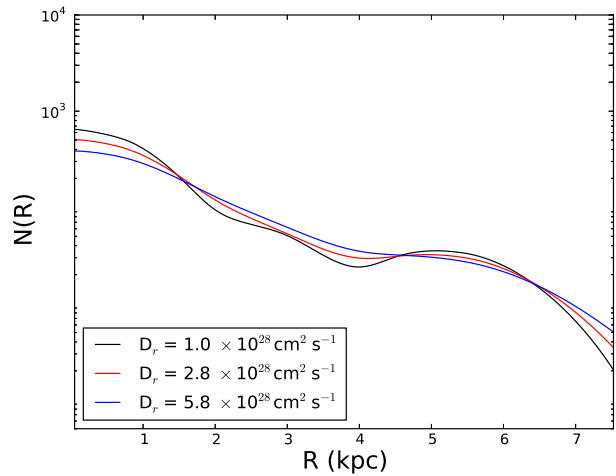}
\end{center}
\caption{The radial distribution of CREs at energies corresponding to an emission frequency of 1.4\,GHz at $5\Myr$ with diffusion using different diffusion coefficients and a continuous injection of CREs. $N(R)$ is in arbitrary units.}
\label{diffusioncontinuousinjectiondifferentd}
\end{figure}

Fig.~\ref{diffusioncontinuousinjectiondifferentd} shows the effect that varying the diffusion coefficient has on the CRE radial distribution after $5\Myr$. Here, due to the continuous injection of CREs, the interarm and arm contrast is much more identifiable than in the single injection case. This contrast is still slightly apparent even for the larger diffusion coefficient $D_{r} = 5.8 \Dunits$.

\begin{figure}[h!]
\begin{center}
\includegraphics[scale=0.45]{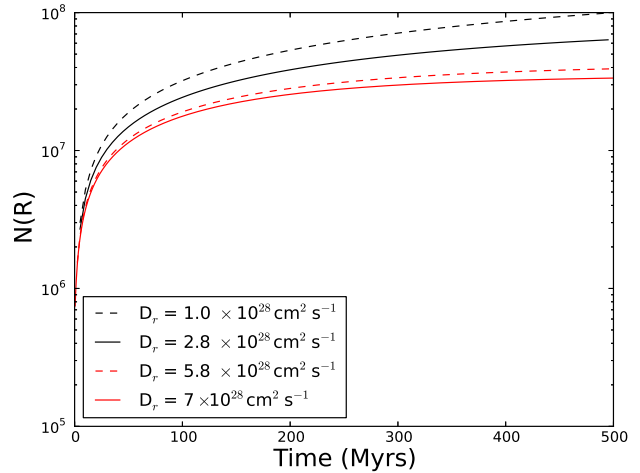}
\end{center}
\caption{The evolution in time of the total number of CREs in M\,51 over the whole galaxy and energies with continuous injection for various diffusion coefficients.}
\label{CREvolutionwithtimedifferentd}
\end{figure}

Fig.~\ref{CREvolutionwithtimedifferentd} shows the evolution of the total CRE population over the whole galaxy and energies with a continuous injection of CREs, showing a sharp increase for the first few Myrs. This increase in population slows as the CREs have time to diffuse through the galaxy and leave at $r\mathrm{_{max}}$. Naturally, the greater the diffusion coefficient, the slower the increase in $N$. For $D_{r}=7\Dunits$, $N(t)$ approaches a constant after about $0.5\Gyr$, as the number of CRE injected at each timestep equals the number leaving the galaxy at $r_{max}=15.05\kpc$.


\section{Diffusion of CREs with energy losses  \label{section5}}

Now we will include the energy losses due to synchrotron emission and inverse Compton scattering. Since these processes depend on $E^2$ they will produce variations in the distribution of $N(r, E, t)$ in time, position and energy. The balance between diffusivity, energy losses and the location of sources determines the distribution of CREs at any given time.

The model CRE source distribution, $Q(r)$, is derived from the radially averaged star formation rate and contains no time-dependence. By time-stepping Eq.~(\ref{eq:1Dnodim}) until a steady solution is reached (i.e. until $\partial N/\partial t = 0$), we are assuming that the observed radially averaged CRE distribution in M\,51 is also in a steady state.

\begin{figure}[h!]
\begin{center}
\includegraphics[scale=0.45]{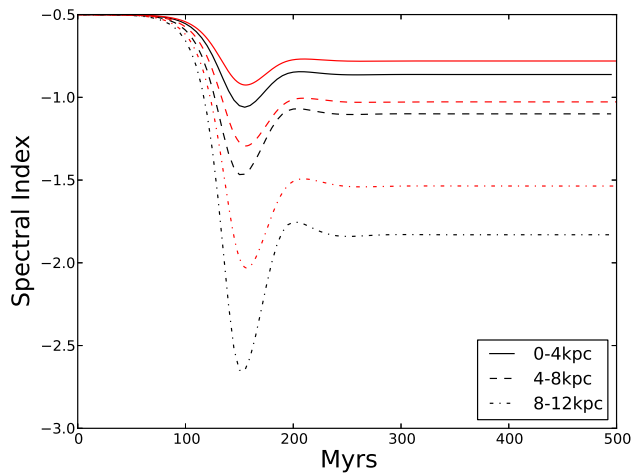}
\end{center}
\caption{The evolution of the spectral index (1.4\,GHz-151\,MHz) at different radial ranges for a model with $D_{r} = 2.8 \Dunits$ (black), $6.6 \Dunits$ (red) and $B=10\uG$.}
\label{fixedspectrumtime}
\end{figure}

Fig.~\ref{fixedspectrumtime} shows that the spectral index, for the case of a constant magnetic field $B=10\uG$, reaches a steady state throughout the model galaxy in a timescale of about $250\Myr$. Initially, the spectral index is constant, as the injection term $Q(r)$ is dominant and the number of CREs in the galaxy builds up (see Fig.~\ref{CREvolutionwithtimedifferentd}). Cosmic ray electrons with energy $E_0$ lose half their energy in the time \citep{Berezinskii:1990,Kardashev1962}
\[
\tau = \frac{1}{\beta E_0}
\]
where $\beta$ is defined in Eq.~(\ref{eq:Eloss}). Thus energy losses for the CREs that emit at $1.4\GHz$, which have the energy of about $3\Gev$ (Eq.~(\ref{eq:nuEnergy}), are significant after $\tau\sim 100\Myr$ and the spectral index steepens significantly after this time. The interplay between CRE injection, diffusive propagation and energy losses then results in variations in the spectral index. An equilibrium is not established until the CREs that originate in the inner galaxy, $r\le 1\kpc$ where the source term is by far the strongest, reach the outer galaxy, which takes a time $\tau\sim L^2/D_r \sim 250\Myr$ for diffusion over $10\kpc$. In the rest of the paper we will only show model results for steady states.

\subsection{Model with energy losses and constant magnetic field strength}

\begin{figure}[h!]
\begin{center}
\includegraphics[scale=0.45]{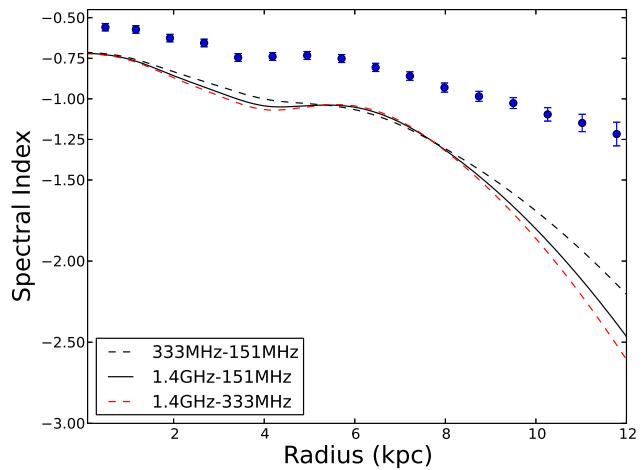}
\end{center}
\caption{Modelled spectral indices between three different frequencies (1.4\,GHz, 333\,MHz and 151\,MHz) from the fixed magnetic field model with $B=10\uG$ and $D_{r} = 2.8 \Dunits$. The observed spectral index between $151\MHz$ and $1.4\GHz$ is also shown.}
\label{fixedballspectrum}
\end{figure}

Initially, we use a constant magnetic field strength to calculate the energy loss rate due to synchrotron emission in Eq.~(\ref{eq:Eloss}), setting $B=10\uG$. Free-free absorption has been neglected.
Fig.~\ref{fixedballspectrum} shows the modelled non-thermal spectral indices between the different frequencies $1.4\GHz$, $333\MHz$ and $151\MHz$. The spectral index was calculated for each pair of frequencies.
The difference seen between the computed synchrotron spectral indices is quite subtle with the largest differences occurring in the extended disk and interarm regions: i.e. in regions where the injection of CREs is low. In these regions, the $1.4\GHz$--$333\MHz$ spectrum is naturally the steepest, with the CREs emitting at 1.4\,GHz losing their energy most rapidly. The flattest spectrum is between $333\MHz$--$151\MHz$, where energy losses are lower than at 1.4\,GHz.

Increasing $Q(r)$, the injection rate of CREs, does not alter the steady-state spectral index of the galaxy. It does increase the time needed for the model to stabilise, as the time that the injection term is dominant becomes  longer making the rapid change in spectral index seen in Fig.~\ref{fixedspectrumtime} when energy losses become important occur later.

\subsection{Model with energy losses and varying magnetic field strength}
\label{subsec:varyB}

Now we include a radially varying magnetic field strength in our model, using the magnetic field profile shown Fig~\ref{bfieldsnrplot}(left). It should be noted that for this profile the magnetic field strength does not drop to a value of $10\uG$ until a radius of $11.5\kpc$. Therefore, we expect the CREs to lose their energy faster and have a shorter diffusion length than in the case of the model with $B=10\uG$.

\begin{figure}[h!]
\begin{center}
\includegraphics[scale=0.45]{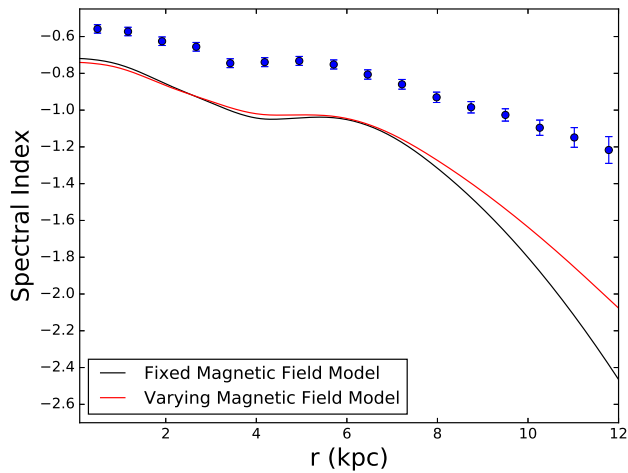}
\end{center}
\caption{Comparison of the synchrotron spectral index $151\MHz$--$1.4\GHz$ between the fixed and variable magnetic field models. In both modelling runs, $D_{r} = 2.8 \Dunits$. The observed spectral index between these frequencies is shown by the blue points.}
\label{comparefixedandvariable}
\end{figure}

Fig.~\ref{comparefixedandvariable} shows the different spectral indices between $151\MHz$--$1.4\GHz$ obtained using the fixed and varying magnetic field profiles. Perhaps surprisingly, both models are similar up to $r=7\kpc$, despite the magnetic field, which enters the synchrotron energy loss term to the second power, being twice as strong for $r\le 1\kpc$ and systematically stronger than $10\uG$ throughout this region. It is only at $7\kpc$ that the spectra diverge. This shows that the the injection of CREs plays a more important role in shaping the spectrum than the radial distribution of the magnetic field strength at radii where CREs are being created.  The main divergence between the two models is only seen in areas where no CREs are being injected, i.e the outer disk.

\begin{figure}[h!]
\begin{center}
\includegraphics[scale=0.45]{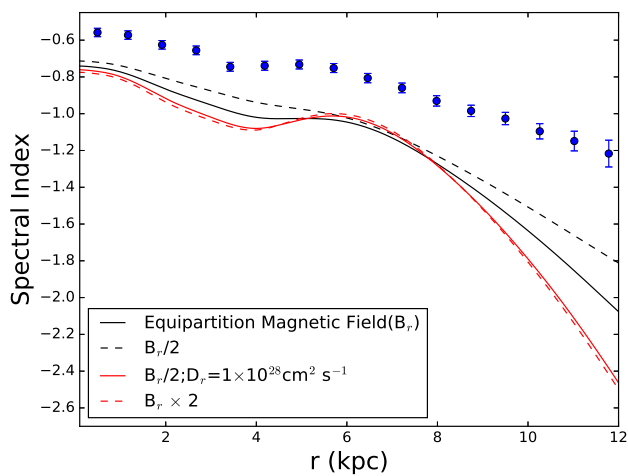}
\end{center}
\caption{The $151$--$1400\MHz$ spectral index for different $B(r)$ profiles. These models use $D_{r}=2.8\Dunits$ unless otherwise stated in the legend. The observed spectral index is shown by blue points with error bars.}
\label{changingbanddif}
\end{figure}

The magnetic field strength profile $B(r)$ of M\,51 that we are utilising has been determined from the synchrotron emission at 151\,MHz by assuming equipartition between the energy densities of cosmic rays and magnetic field, using the revised formula of \cite{2005AN....326..414B}. While the equipartition assumption tends to give plausible magnetic field strength estimates when it can be compared to other methods, such as the interpretation of Faraday rotation measures \citep[e.g.][]{Fletcher:2004}, there is no physical requirement for equipartition, but the interaction between CREs and magnetic fields leads to an energy balance beyond some minimum scale in space and time \citep{Duric:1990}.  Indeed there is evidence that the assumption does not hold on sub-kpc scales in the Milky Way \citep{Stepanov:2014}. That equipartition does not hold on small scales is the direct effect of CR diffusion, as was shown for M31 by \citep{2013MNRAS.435.1598B}. On the other hand, we have just seen that moving from a model with $B=10\uG$ at all radii to a model with a strong dependence of $B$ on radius, had little effect on the spectral index profile. Next, we shall check how variations of a factor-of-two in the estimated equipartition magnetic field strength profile change the spectral index. A factor of two in difference in the equipartition field strength does not necessarily mean a deviation from equipartition, but could just be due to the uncertainties in the proton to electron ratio and the pathlength.

Fig.~\ref{changingbanddif} shows spectral index profiles for different estimated magnetic field strengths. The synchrotron spectrum flattens overall for a weaker magnetic field strength since the high energy CREs lose their energy more slowly. This effect is especially pronounced in regions where the CRE injection is lower, such as the interarm region and extended disk. There is a clear degeneracy between changing the field strength and changing the diffusion coefficient: increasing the diffusion coefficient has a similar effect to decreasing the magnetic strength, as in both cases the CREs can travel further before they lose significant energy. Decreasing the magnetic field strength towards zero shows the synchrotron spectrum to flatten to its injection spectrum as expected.
 
\begin{figure}[h!]
\begin{center}
\includegraphics[scale=0.45]{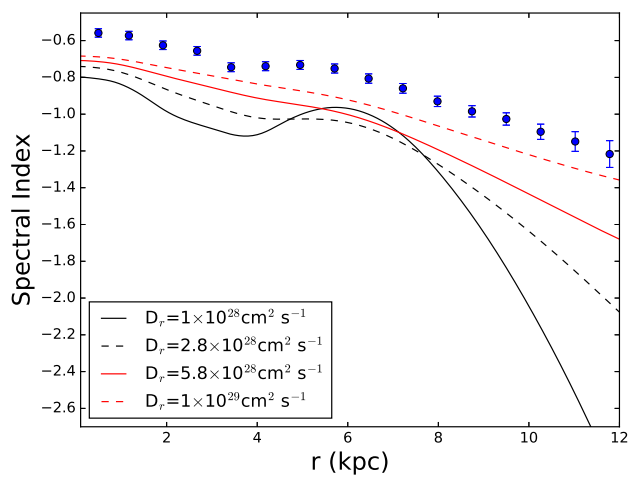}
\end{center}
\caption{The $151$--$1400\MHz$ spectral index for different diffusion coefficients using the radial profile of magnetic field strength shown in Fig.~\ref{bfieldsnrplot}. The observed spectral index between these frequencies is shown by the blue points, with error bars.}
\label{changingbdifonly}
\end{figure}

Fig.~\ref{changingbdifonly} shows how changes in the diffusion coefficient affect the spectral index. Two types of behaviour are clearly shown: a higher diffusion coefficient results in a flatter spectrum and also in a smoother radial profile of the spectral index. Both of these effects have the same origin, in the differing ratio of CRE propagation to injection. Low $D_{r}$ means that the CREs tend to lose their energy, and thus change their spectral distribution, closer to their source than in the case of a high $D_{r}$. Even for the highest diffusivity, the modelled spectrum is systematically steeper than the observed spectrum.
 


So far the synchrotron spectral indices produced by our model have been systematically too steep compared to the observations. We saw in Fig.~\ref{changingbanddif} that reducing the magnetic field strength by a factor of two does not flatten the spectrum sufficiently neither does increasing the diffusion coefficient to $D_{r}=1\Dunitshigh$, a very high diffusivity, as seen in Fig.~\ref{changingbdifonly}. We shall now address this inconsistency.

\section{Vertical escape of CREs from M\,51 \label{section6}}

All of the models we have considered so far have produced spectra that are steeper at all radii than those observed. This is because the CREs in the model are losing too much energy before leaving the system, either via diffusion across the outer boundary at $r=15.05\kpc$ or by losing sufficient energy to be removed from our lowest energy bin at $E=10\Mev$. In all of these models we have only considered diffusion in the radial direction. However, it is inevitable that there will also be diffusion in the vertical direction and that this will lead to CREs being lost to the galaxy. Since the scale height of the galaxy is smaller than its radial extent, the escape time due to diffusion in the vertical direction will be shorter than the time taken to diffuse to the outer radial boundary, if the diffusivity is the same in each direction.

The timescale for CREs to diffuse one magnetic field scale height $h$ from the disc mid-plane is
\begin{equation}
\tau = \frac{h^2}{D_z},
\label{eq:esc}
\end{equation} 
where $D_z$ is the diffusion coefficient in the vertical direction. A similar relation governs the time for CREs to diffuse from the galaxy centre to the outer radius, with $r$ and $D_r$ replacing $h$ and $D_z$ respectively. Thus the time scale for vertical losses is much shorter than that for radial losses, by the factor $r^2/h^2\approx 15^2/5^2 = 9$, if the diffusion is isotropic with $D_r=D_z$. Models by \cite{1977A&A....61...59S} and \cite{2010ApJ...717....1L} include a simple escape term, controlled by an escape time $\tau_{\mathrm{esc}}$.

\subsection{Vertical escape of CREs by diffusion}

\begin{figure}[h!]
\begin{center}
\includegraphics[scale=0.45]{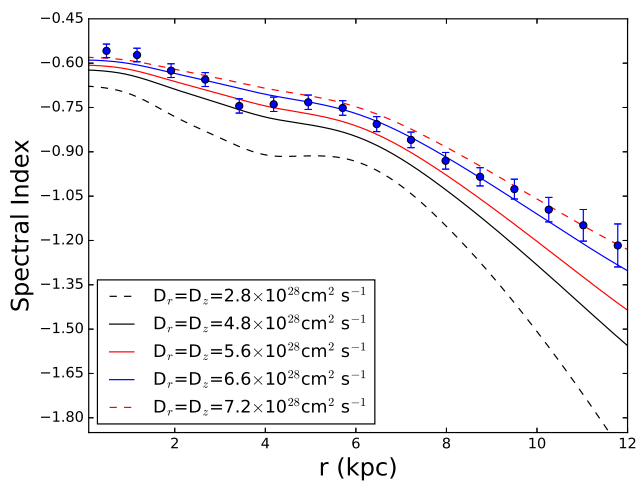}
\end{center}
\caption{Spectral index of M\,51 for different diffusion coefficients both in radial and vertical directions. }
\label{comparedz}
\end{figure}

In order to include vertical diffusion in our model we use Eq.~(\ref{eq:esc}) to define the escape time $\tau$ in the last term on the right hand side of Eq.~(\ref{eq:1Dnodim}), set $D_z=D_r$ unless otherwise specified and adopt the best available model for the magnetic field scale heights of M\,51 that we have. \cite{1997A&A...318..700B} estimated the synchrotron scale height for M\,51 at $\lambda$20.5cm as a function of radius. The values found correspond to magnetic field scale heights of $h=3.2\kpc$ at radii $3<r<6\kpc$ rising to $h=8.8\kpc$ for $12<r<15\kpc$. Our model here uses the magnetic field scale height. The result is shown in Fig~\ref{comparedz}.

When $D_r = D_{z} = 6.6\pm0.2\Dunits$ the modelled spectral index produces a very good match to the observational data, with the regions at $2<r<3\kpc$ and $5<r<12\kpc$ being within the error bars. The interarm-dominated ring at 3\,kpc, has a steeper observed spectrum than neighbouring radii, which is not matched by the model spectral indices. Since the source function $Q(r)$, derived from the observed star formation rate, is weaker at this radius, this may indicate that there should be a dependence of  the diffusivity on the star formation rate. Additionally, there is still a thermal contribution in the 1.4 GHz data which would make the observed spectra flatter than the modelled synchrotron-only spectra, but this effect would be weakest where the thermal emission is weakest, in the interarm regions.  The non thermal spectrum for values D$_{z} < 5.6\Dunits$  are seen to be too steep compared to the observational data.

For $D_{z} = 6.6 \Dunits$, we obtain an escape time  i.e. the timescale on which the CREs leave the galaxy of 11\,Myrs at the centre of M\,51 and 88\,Myrs at a radius of 12\,kpc. 
For comparison, based on measurements from the ULYSSES high energy telescope, \cite{1998ApJ...501L..59C} found, from measurements of $^{10}$Be (an ideal CR clock), that the confinement time of cosmic rays in the Milky Way was found to be is $26^{+4}_{-5}$ Myrs. 

\begin{figure}[h!]
\begin{center}
\includegraphics[scale=0.45]{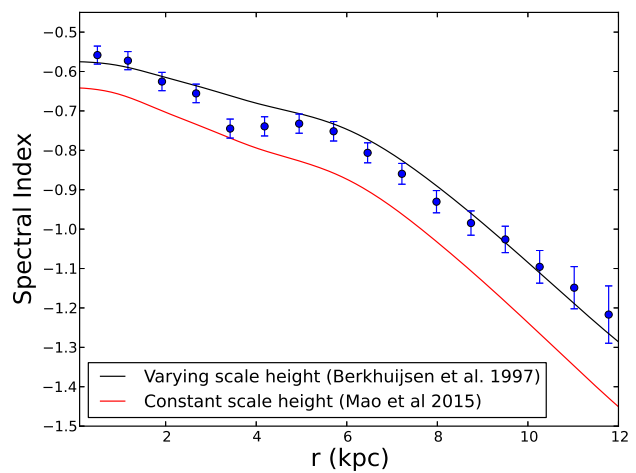}
\end{center}
\caption{The comparison of the radial spectral index of M\,51 using a constant and varying scale height with radius.}
\label{comparedzconstantscale}
\end{figure}

The variation of the scale height with radius has a more pronounced effect on the model's spectral index distribution than the variations in magnetic field strength discussed in Sect.~\ref{subsec:varyB}.
Fig.~\ref{comparedzconstantscale} shows the difference between the spectra obtained from a model using a constant magnetic scale height of $5.6\kpc$ from \cite{2015ApJ...800...92M}. The difference in spectral index is not constant, varying from 0.08 in the central region of M\,51 to 0.16 at around 11\,kpc.

Compared to other galaxies, \cite{2013MNRAS.435.1598B}  showed that in M 31 most of the CREs are confined to a thin non thermal disk
with an exponential scale height of 0.3 $\sim$ 0.4 kpc at 1.4 GHz, whereas in M 33 many CREs move out of the thin disk into a thick
disk / halo with scale height of $\sim$ 2 kpc at 1.4 GHz. A thick non thermal disk or halo in M 33  is consistent with the
higher SFR in M 33 and the existence of a vertical magnetic field component.

\subsection{Comparison between analytical and numerical models}

\citet{1977A&A....61...59S} compared analytic models for the CRE propagation in M\,51 to spectral index measurements between $610\MHz$--$1.4\GHz$, obtained with the Westerbork Synthesis Radio Telescope in the 1970s. We will compare the most realistic of Segalovitz's models, which takes the source distribution to be a step-function in radius and fixes the lifetime for CREs in the galaxy as $30\Myr$, with our numerical model. We use the same parameters as Segalovitz: $D=1\Dunitshigh$, $B=5\uG$ and the energy spectral index of the CRE sources is $p=2.18$.

\begin{figure}[h!]
\begin{center}
\includegraphics[scale=0.45]{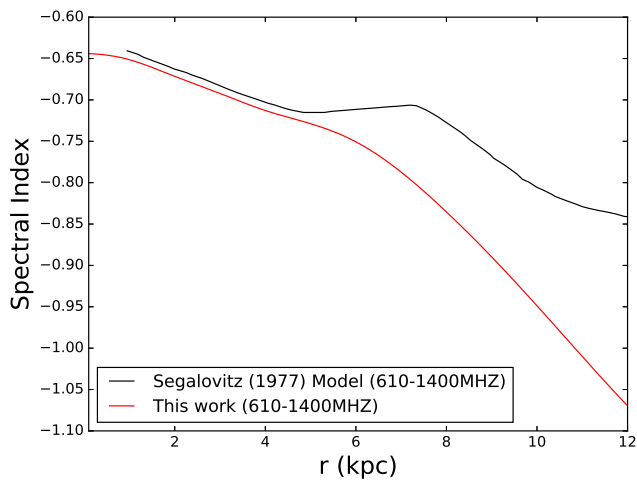}
\end{center}
\caption{Comparison between the analytical model of  \cite{1977A&A....61...59S} and the numerical model developed from this work, using the parameters described in the text.}
\label{segalovitzplot}
\end{figure}

Fig.~\ref{segalovitzplot} shows the predicted $610\MHz$--$1.4\GHz$ spectral index for the analytical model of \cite{1977A&A....61...59S} and our numerical model. Up to 5\,kpc, both models agree very well. At radii greater than 6\,kpc, the models diverge with the analytical model flattening slightly at 7\,kpc and gently decreasing to $\alpha=-0.8$ at 12\,kpc. Our numerical model shows that $\alpha$ does not flatten but steepens at a greater rate with $\alpha>-1.0$ at 12\,kpc. Thus, the simple analytic model, which can be solved exactly using a Green's function, performs well in the part of the disc where the CRE sources are significant but fails to properly account for CRE energy losses where the sources are weak.

\subsection{Energy dependent diffusivity \label{section7}}


There is debate about the relevance of the energy dependence of the diffusion coefficient  \citep{2006JPhCS..47..113P} at low energies, which correspond to low radio frequencies.


Firstly, the energy at which the energy dependence in diffusion becomes relevant is usually at 1\,GeV according to \cite{2007ARNPS..57..285S} and \cite{1992ARA&A..30..575C}.
Some studies, when investigating the FIR--Radio correlation \citep{2008ApJ...678..828M, 2013A&A...552A..19T} routinely use this energy value.
Others claim that the energy dependence in diffusivity does not become relevant until 4\,GeV \citep{1997ICRC....4..245V}. 
For energy dependent diffusion,  it is found from observations from Voyager-2 of beryllium isotope ratios at the Solar Circle that the confinement timescale 
for particles $\geq$ 3\,GeV is dependent on energy \citep{2003JGRA..108.1355W}. The dependence of the CRE lifetime in the Galaxy, $t_\mathrm{diff}$, on CRE energy $E$,
\begin{equation}
t_{\mathrm{diff}}(E) = 26\, \mathrm{Myr}\left(\frac{E}{3 \mathrm{GeV}}\right)^{-1/2},
\end{equation}
is used in models by \cite{2010ApJ...717....1L}.
This 3\,GeV condition they use allows them to model the radio far infrared correlation from normal spirals to dense starburst and were able to reproduce a linear correlation consistent with both local and integrated Galactic constraints on the
energy in CR protons.
Finally, the review of \cite{2012JCAP...01..010B} indicates that the energy dependence begins at 3\,GeV.

Assuming diffusion is isotropic, we can study the effect of a varying diffusivity for electrons of different energies by writing

\begin{equation}
D(E) = \begin{cases} D_0 & \quad \text{if} E\leq E_0 \\ D_0E^{\kappa} & \quad \text{if } E\geq E_0\\ \end{cases}
\end{equation}
where $\kappa$ determines how strong the variation in diffusivity is and $E_0$ sets the energy at which energy dependent diffusion becomes important. The diffusive escape time using this model is
\begin{equation}
t_{\mathrm{diff}}(E) = \frac{h^2}{D_{z}}\left(\frac{E}{E_0}\right)^{-\kappa},
\label{eq:energyDiff}
\end{equation}
where $h$ is the scale height of the galaxy.

The value of $\kappa$ is uncertain and depends upon the theoretical model for diffusion. Using quasi-linear theory and a magnetic field with a Kolmogorov-like power spectrum of magnetic fluctuations, \citep{2002cra..book.....S} obtained $\kappa\approx 0.3$. Utilising a full Bayesian parameter estimation for GALPROP, \cite{Trotta:2011} found that  $\kappa=0.3$ obtained the best global fit with second order Fermi acceleration as the source of cosmic rays at lower energies. This allowed a better reproduction of the Boron/Carbon ratio. Using a weakly non-linear theory, \cite{2005ApJ...626L..97S} found $\kappa=0.5$--$0.6$ and were able to successfully predict the abundance ratio of secondary to primary cosmic ray nuclei in the Milky Way. However, the strong diffusion resulting from $\kappa=0.5$--$0.6$ leads to a large cosmic ray anisotropy \citep{2006AdSpR..37.1909P}.

\begin{figure}[h!]
\begin{center}
\includegraphics[scale=0.45]{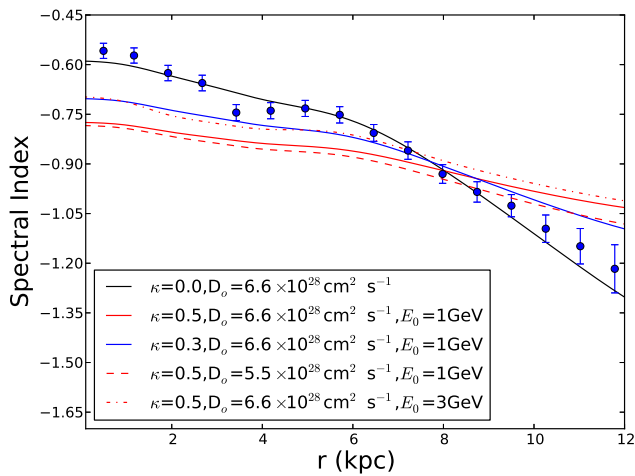}
\end{center}
\caption{Comparison between energy dependent diffusion, with different energy dependence, diffusion coefficient and starting energy in Eq.~(\ref{eq:energyDiff}) .$\gamma_0$ is always set to 2.0. }
\label{energydependencediffusionkappa5}
\end{figure}

In order to investigate these questions, we implemented the above energy dependent diffusion equations into our model and varied several parameters such as $\kappa$, E$_0$ and D$_0$. The injection was kept at $\gamma_0$=2.0.
Fig.~\ref{energydependencediffusionkappa5} shows the result of these tests; that the implementation of the energy dependence in the diffusion coefficient makes the spectrum too steep. Additionally, the change of spectral index with radii is too small.  The most inaccurate values are seen with $\kappa$ = 0.5 and E$_0$ = 1 GeV. We can say confidently that the energy dependence in the diffusion coefficient does not play a part at such energies.

For E$_0$ = 3 GeV, we find again that the spectrum is too steep compared to observations, but the change of spectral index with radii is closer to what is observed. This gives us the indication that the energy dependence of the diffusion coefficient starts towards higher energies and is not important for energies in the range $0.01$--$3\Gev$.

If the energy dependence in the diffusion coefficient becomes relevant at energies above 4\,GeV,  one would expect that this would effect only the CRE population emitting  above 3-4\,GHz.
The energy dependence of diffusion, while not directly operating at the energies where the CREs are emitting, can still affect the spectral index at lower energies. This is due to the fact that the CREs at higher energies are able to leave the galaxy faster and therefore a smaller CRE population are able to lose their energies and cascade into 1.4\,GHz energy emitting range. This would decrease the flux emitted at 1.4\,GHz and therefore steepen the nonthermal spectrum.
Alternatively, if energy dependent diffusion is not present, the CRE population at higher energies is now greater as they leave the galaxy at the same rate as the lower energy CREs. These CREs then pass down to 1.4\,GHz energy emitting range and increase the flux thereby flatten the nonthermal spectrum. This effect is also seen in our model.

\section{Radio continuum scale lengths}
\label{sec:ScaleLength}
\begin{figure}[h!]
\begin{center}
\includegraphics[scale=0.45]{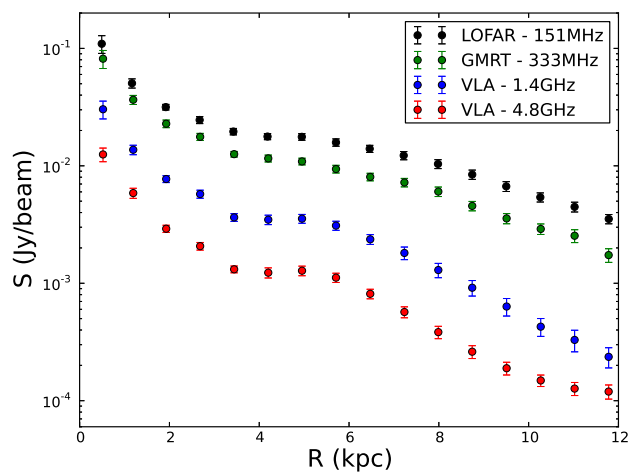}
\end{center}
\caption[The radial profile of the radio continuum for M\,51 at several frequencies.]{The radial profile of the radio continuum for M\,51 at several frequencies.}
\label{allfrequenciesradial}
\end{figure}

From the observed radio continuum maps, the radial profile of M\,51 at four frequencies was calculated in concentric rings, with the position angle of the major axis and the
inclination of the galaxy taken into account using the values from Table~\ref{physicalpara}. Data at 151\,MHz \citep{2014A&A...568A..74M}, 325\,MHz (this work), 1.4\,GHz and 4.8\,GHz \citep{2011MNRAS.412.2396F} were used. Background point sources were removed by fitting Gaussians to them before measuring the radial profile.
Several background point sources located in the disk were also blanked out. Additionally emission from, the companion galaxy, NGC\,5195, was subtracted. The radial profiles are shown in Fig.~\ref{allfrequenciesradial}.
The difference between radii dominated by arm and interam regions is  more apparent at the higher frequencies but becomes smeared out with decreasing frequency. This is due to the faster energy losses of high energy CREs which results in them only diffusing a short distance from their source.

\begin{figure}[h!]
\begin{center}
\includegraphics[scale=0.45]{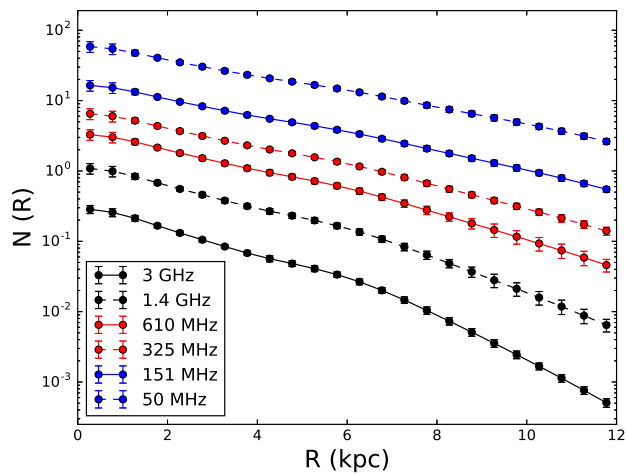}
\end{center}
\caption[The radial profile of the radio continuum for M\,51 at several frequencies.]{The radial profile of the CRE population for several frequencies, obtained from our model.}
\label{allfrequenciesradialmodel}
\end{figure}

The modelled radial distribution of the CRE populations at different frequencies can be seen in Fig.~\ref{allfrequenciesradialmodel}, obtained with $D_{r}=D_{z} = 6.6 \Dunits$.
The observed radio continuum profile changes slope at approximately 5\,kpc. This effect was also seen by \citet{2014A&A...568A..74M}, the presence of NGC\,5195 created a much more obvious change. Such a change in slope in the observed radio continuum is also observed in M\,101 \citep{Berk2016}.
A similar radial distribution of emissivity is also seen in the modelled profiles but the change in slope occurs at a slightly larger radius of 6\,kpc. The change in radial profile becomes less obvious at lower frequencies, both in the model and observations.
This break coincides with a break seen in the star formation and magnetic field profiles shown in Fig.~\ref{bfieldsnrplot}.

We fitted exponential profiles of the form $I(r)=I_0\exp{-(r/l)}$ to the radial ranges $1\kpc \le r\le 6\kpc$ (avoiding the galactic nucleus, where there is weak AGN activity) and $6\kpc\le r \le 12\kpc$. The fitted exponential scale lengths are shown in  Table.~\ref{modelscalelengthtable} for both observations and our model. 

The scale lengths increase with decreasing frequency, which is what one would expect as the CREs at low frequencies lose their energies at a slower rate and therefore can travel further, thus increasing the scale length.
Both the models and observations show that the scale length in the outer disk is significantly less that the inner disk. This is due to the fact that there are few sources of CREs in the outer disk (in our model there is no injection of CREs at large radii at all).  The smaller scale lengths in the outer disk of M\,51 are due to propagation effects: the absence of sources at $r>8\kpc$ (in the model $Q=0$ for $r>8\kpc$, in the galaxy $Q\rightarrow 0$ at large $r$) means that the emitting CREs have arrived from the inner galaxy and have already lost energy on the journey.  Our model consistently underestimates the scale length at all frequencies for both the inner and outer disk. This is likely because there are some CREs sources at larger radii.

 \begin{table*}
\centering
\begin{tabular}{c c c c c}
\hline\hline
Frequency & Modelled scale length & Observed scale length & Modelled scale length & Observed scale length\\
& (R $<$6\,kpc) & (R $<$6\,kpc) & (R $>$6\,kpc) & (R $>$6\,kpc) \\
MHz & kpc & kpc & kpc & kpc \\
\hline

4800 & - & 3.1$\pm$0.4 & - & 2.2$\pm$0.1 \\

3000 & 2.46$\pm$0.01 & - & 1.38$\pm$0.01 & - \\

1400 & 2.82$\pm$0.05 &  3.5$\pm$0.5  & 1.80$\pm$0.02 & 2.3$\pm$0.1 \\

610 & 3.17$\pm$0.05 & - & 2.28$\pm$0.02 & - \\

325 & 3.38$\pm$0.05 & 3.8$\pm$0.4 & 2.62$\pm$0.02 &  3.3$\pm$0.1   \\

151 &  3.64$\pm$0.08 &  4.7$\pm$0.6   & 3.07$\pm$0.03 &  3.6$\pm$0.2   \\

50 & 3.85$\pm$0.07 & - & 3.47$\pm$0.04 & - \\

\hline
\end{tabular}
\caption{Exponential scale lengths of the model CRE population radial profiles.}
\label{modelscalelengthtable}
\end{table*}

\begin{figure}[h!]
\centering
\includegraphics[scale=0.45]{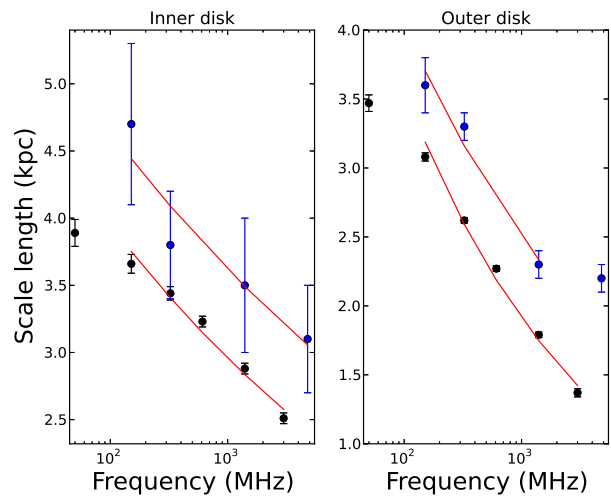}
\caption{Measured continuum and modelled CRE population scalelengths, with fitted power laws shown in red. The blue and black data points signify the observed and modelled scale lengths respectively.}
\label{frequenciesagainstscalelength}
\end{figure}

The observed and modelled scale lengths are plotted against frequency in Fig.~\ref{frequenciesagainstscalelength}. Although the scale lengths derived from the model are systematically shorter than those observed, both model and observations show a similar dependence of $l$ on $\nu$.

We have fitted a power law to the scale lengths against frequency for both model and observed values. The results of these fits are shown in Table~\ref{fittedscalelengthtable}.

We can derive the expected dependence of $l$ on $\nu$ as follows. The diffusion length is
\begin{equation}
 l \propto \sqrt{D \cdot t_{syn}}
\end{equation}
with the synchrotron lifetime given by
\begin{equation}
 t_{syn} \propto B^{-3/2} \nu^{-1/2}
\end{equation}
thus
\begin{equation}
 l \propto \nu^{-1/4} B^{-3/4}.
\end{equation}

Both our model and  the observed data show this dependence of scale length on frequency in the outer disk up to frequencies of 1.4\,GHz. Beyond this frequency, the energy dependence could start becoming important. In the outer galaxy, the CRE sources are either non-existent (the model) or very weak (the galaxy) and so the balance of, diffusion to energy loss rate (approximated by $t_\mathrm{syn}$ above) plays the dominant role in shaping the scale lengths and spectra. In  the inner galaxy the distribution of CRE sources $Q(r)$ must also be taken into account and so the simple estimate of $l\propto \nu^{-1/4}$ derived above no longer applies. 
 
Two exceptions to the systematic decrease of the scale length with frequency are shown in Fig.~\ref{frequenciesagainstscalelength}. The first exception is that the modelled synchrotron emission at $50\MHz$, in both the inner and outer galaxy, has a shorter exponential scale length than expected from the fitted power law. This would mean that a a very extended disk in M51 will be difficult to observe at very low frequencies with the LOFAR LBA. The second exception being that the observed scale length in the outer galaxy at $4.8\GHz$ is longer than expected.

The scale length for the observed continuum at the outer disk for 4.8\,GHz is very similar to 1.4\,GHz. This is not seen in our model and could be due to either errors in the observed data or an incorrect assumption of the model. (The most likely source of observational error is probably in the combination of single dish and interferometric measurements.)

\begin{table}
\centering
\begin{tabular}{c c c}
\hline
 &(R $<$6\,kpc) & (R $>$6\,kpc) \\
\hline
Model & -0.13$\pm$0.01 & -0.26$\pm$0.02 \\
Observed & -0.11$\pm$0.02 & -0.21$\pm$0.03\\
\hline
\end{tabular}
\caption{Fitted power law with frequency against scale length for both observed continuum and modelled CRE population.}
\label{fittedscalelengthtable}
\end{table}

\section{Discussion \label{section8}}


The results from our numerical models show that a diffusion coefficient of approximately $D = 6.6 \pm0.2 \Dunits$  matches observed values of the spectral index between 151\,MHz and 1.4\,GHz and that diffusion alone is sufficient to replicate the observed spectral values.
This value is greater than what is found in \cite{2014A&A...568A..74M}, with $D = 3.3 \Dunitslow$.  While this is smaller compared to the value found in this work, it should be noted that the value found in \cite{2014A&A...568A..74M} is only an approximation, accurate to the order of magnitude of the diffusion coefficient. The value for the diffusion coefficient found in this work using a much more sophisticated model is deemed to be more accurate.
One issue with the estimation from \cite{2014A&A...568A..74M} could have been the underestimating of the diffusion length. The diffusion length was chosen by comparing the correlation between radio and FIR data and a small change 
to the criteria used to select the correlation scale would result in a diffusion length at 151 MHz to be approximately $2\kpc$ which leads to $D = 6.6 \Dunits$, in agreement with our present results.

We find that the value of the diffusion coefficient found in this work, while comparable is larger to what is found for the range of perpendicular diffusion values \cite{2013ApJ...764...37B}.  We would anticipate that the diffusion coefficient parallel to the magnetic field lines would dominate in our work, both in the disk and halo and therefore we would expect a larger diffusion coefficient. Additionally, our value for the diffusion coefficient parallel to the magnetic field lines in M\,51 is smaller than what is observed in NGC\,253 \citep{2009A&A...494..563H}, which is expected as NGC\,253 is a powerful starburst galaxy.

 The observed radio continuum emission that we have used to compute spectra and scale lengths still contains free-free emission. This thermal component at $151\MHz$ is expected to be very low and will not affect the results at the level of accuracy at which we can work. But at $1.4\GHz$, the thermal fraction could be of order 10\% on average, higher in the inner galaxy where the star formation rate is higher and lower in the outer galaxy.
This would lead to flatter observed spectra, especially in the star forming region of the disc $r\lesssim 6\kpc$.

While we exclude energy dependent diffusion at energies less than $E_0=3\Gev$ (Sect.~\ref{section6}), further investigation is needed for the energy ranges greater than $4\Gev$. This would help us determine whether a strong ($\kappa = 0.5\sim0.6$) or weak ($\kappa = 0.3$) scaling of diffusivity with CRE energy holds true.
However an accurate thermal separation at frequencies greater than $1.4\GHz$ is necessary to derive synchrotron spectra that are reliable enough to test energy dependent diffusion at these energies.
At even lower frequencies (LBA frequencies), free-free absorption will become especially relevant and would need to be included into this model as it has been neglected in this work.

Not only do relativistic electrons leave via diffusion but in starburst galaxies, large scale galactic winds can advect CREs out of the disk into the halo with a short timescale compared to diffusion observed in M82 \citep{2013A&A...555A..23A} and NGC253 \citep{2009A&A...494..563H}.  There is no observational evidence for a global outflow or a galactic-scale wind in M\,51 \citep{2015ApJ...800...92M}.
Our results show that isotropic diffusion in the vertical and radial direction fits the observed spectrum most accurately and there is no need for a wind.
If CRE diffusivity is much lower in the z direction, then a wind could possibly exist.
It is shown observationally \citep{1995ApJ...444..119D,2009A&A...494..563H} that the diffusion coefficient  in the z direction is several times lower than the radial component.
\cite{2016MNRAS.tmp..146H}, using 1-D cosmic ray electron transport models and Australia Telescope Compact Array observations of two edge-on galaxies, found that the cosmic ray transport in the halo of NGC\,7090 is advection dominated with a velocity of approximately 150 km s$^{-1}$, while NGC\,7462 is diffusion dominated with a diffusion coefficient of $D = 3.0 \times 10^{28} E^{0.5}_{GeV}$.
In the case of M\,51, an upper limit for such a wind if $D_z = 0$ would be 100 km s$^{-1}$ for the centre of the galaxy. In the case that $D_z < D_r$ but not equal to zero, then a combination of outflow and vertical diffusion may be necessary
What is certain is that a significant amount of CREs are needed to leave the disk into the halo and thus a significant vertical magnetic field needs to exist in M\,51 to produce the observable spectrum.

Improvements could be made in the radial distribution of the CRE injection source, as we have assumed no injection of CREs beyond 7.5\,kpc. \cite{2006ApJ...651L.101T} determined the star formation rate (SFR) for regions with faint HI and infrared emission  up to $13\kpc$ from the centre of the galaxy. However, no radially averaged SFR was computed and it was found that the star formation efficiency in the extended disk was several magnitudes lower than the star forming region. This could be important to why the modelled scale length for the outer disk is shorter than what is observed, as including injection in this region would increase the scale length. 
In case of CRE diffusion, equipartition is not valid in the outer disk, i.e. the field strength is underestimated and so are the synchrotron losses. This would make the model spectral index somewhat steeper in the outer parts.
Additionally, inaccuracies on the scale height with radius could also play a role in these inaccuracies, especially if D$_z$ has a radial dependence.

In this work we have assumed that the diffusive transport of CREs is described by a scalar diffusion coefficient whereas a diffusion tensor is required to take into account that the parallel and perpendicular diffusion are different \citep{2010ApJ...725.2117S} \citep{1966ApJ...146..480J}. In addition to this, anisotropic diffusion is also seen to be essential for CR-driven magnetic dynamo action in galaxies \citep{2009A&A...498..335H}. Including a moderate anisotropy for diffusion has given more consistent results with recent estimates of the local interstellar proton spectrum compared to normal isotropic diffusion \citep{2012A&A...547A.120E}. As this model has only one spatial dimension , we are only implementing a one halo model. A two halo model would be more physical consistent \citep{2015arXiv150905775T} as the increased turbulence in the disk, compared with that of higher latitudes would produce a spatial change in the CRE diffusion properties.

\subsection{Predictions of the spectral index of M\,51 at LBA frequencies}

\begin{figure}[h!]
\begin{center}
\includegraphics[scale=0.45]{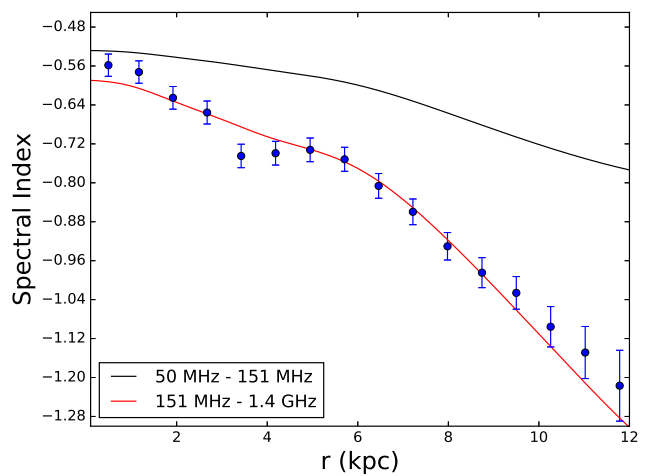}
\end{center}
\caption{Prediction of the radial variation of the spectral index between $50\MHz$--$151\MHz$ (black line), plotted with the observed points with error bars and modelled spectra between $151\MHz$--$1.4\GHz$.}
\label{lbapredict}
\end{figure}

LOFAR \citep{2013A&A...556A...2V} consists of two different types of antenna, LBA (Low Band Antenna) and HBA (High Band Antenna),operating at frequenices of 10-80 MHz and 120-240 MHz respectively. Fig.~\ref{lbapredict} shows the variation of the spectral index between $50\MHz$--$151\MHz$ with radius predicted by our model, using $D=6.6\Dunits$. The spectrum is expected to be significantly flatter at all radii, due to the slower energy losses of CREs at these energies. However, this does not take into account absorption of the synchrotron emission by thermal plasma, which can be important at low frequencies.

\cite{2014A&A...568A..74M} observed a spectral index of $-0.47 \leq \alpha \leq -0.52$ between $151\MHz$--$1.4\GHz$ in the
central region and inner spiral arms of M\,51, which is flatter than expected from CRE acceleration models. If this flattening is indeed caused by thermal free-free absorption,
then we would expect the  spectral index to be slightly flatter than shown in Fig.~\ref{lbapredict}, especially for $r\lesssim 2\kpc$ where the SFR and thus thermal electron density are high.

LOFAR LBA observations are in progress and the comparison between this prediction and observations will be explored in a further paper (Mulcahy et al. in prep).
Observations of M\,51 from LOFAR LBA would let us probe even further into the extended disk and give us more a more accurate nonthermal spectral index and magnetic field strengths beyond 12\,kpc.

\section{Conclusions \label{section9}}

High-resolution low-frequency observations of nearby galaxies can now be made with new interferometers such as LOFAR. These data are particularly useful for studying the propagation of CREs as they are essentially uncontaminated by thermal radio emission and the wide gap in frequency to readily available $1.4\GHz$ maps allow accurate spectral indices to be determined as a function of position in the galaxy. Thus observations and modelling of external galaxies will be a useful adjunct to studies of CRE propagation in the Milky Way, such as those using the GALPROP code \citep[see discussion in e.g.][]{Strong:2007, Grenier:2015}.

Our main findings are summarised as follows:
\begin{enumerate}
\item Transport by diffusion alone is sufficient to reproduce the observed radial spectral index profile of M\,51. 
\item We find an isotropic diffusion coefficient of  $D = 6.6 \pm0.2 \Dunits$ can fit the observed spectral index of M\,51 at all radii.
\item The escape time of CREs in M\,51 is about $11 \Myr$ in the inner galaxy, increasing to over $88 \Myr$ in the outer galaxy.
\item We are able to replicate the break seen in the observed total intensity radial profiles at approximately 6\,kpc. This is due to a break in the star formation rate, which determines the injection rate for CREs at this radius.
\item Energy dependence of the diffusion coefficient is not needed at the CRE energies in the range $0.01\Gev$--$3\Gev$. Reliable high frequency observations which have been thermally subtracted are essential to investigate if the energy dependence becomes important at higher CRE energies. 
\item We predict that the low frequency non-thermal spectrum between both observing bands of LOFAR should be flatter than $-0.6$ for $r<6\kpc$ and steepen to $-0.7$ at the edge of the galaxy. Any major deviation from this spectrum could signify other physical effects are in play such as free-free absorption. 
\item We are able to replicate the frequency dependence of the observed scale lengths with a dependence of approximately $l \propto \nu^{-1/4}$  and $l \propto \nu^{-1/8}$ for the outer and inner disk, respectively.
\item The radio continuum scale length predicted for frequencies below 100\,MHz, is smaller than expected from the observed and modelled frequency dependence of the scale length and would mean that a very extended disk in M\,51 will be difficult to observe.
\end{enumerate}

Obvious extensions to this work, such as including a time dependent CRE injection in the form of the star formation history of the galaxy, inserting a radial dependence on the diffusion coefficient, using the 2D spectral index map to include arm and interam variations in the model of M\,51 and other face-on galaxies where low frequency radio continuum data is available and using new low-frequency observations with LOFAR to model vertical variations in the synchrotron spectral index in edge-on galaxies, are planned.

\begin{acknowledgements}
We thank the staff of the GMRT that made these observations possible. GMRT is run by the National Centre for Radio Astrophysics of the Tata Institute of Fundamental Research.
This research was performed in the framework of the DFG Research Unit 1254 ``Magnetisation of Interstellar and Intergalactic Media: The Prospects of Low-Frequency Radio Observations''. 
D.D.M is grateful to the University of Newcastle for their support and hospitality. We thank Dr. Aritra Basu for careful reading of the manuscript and useful comments. We thank the anonymous referee for their helpful comments which helped improve the paper. 
 D.D.M thanks J.J. Cheung for the Takoyaki and the reminder who the real Genius is.
LOFAR, designed and constructed by ASTRON, has facilities in several countries, that are owned by various parties (each with their own funding sources), and that are collectively operated by the
International LOFAR Telescope (ILT) foundation under a joint scientific policy. 
D.D.M and A.M.M.S acknowledge support from ERCStG 307215 (LODESTONE). A.F. acknowledges grants from STFC (ST/L005549/1) and the Leverhulme Trust (RPG-097).
\end{acknowledgements}

\bibliographystyle{aa}
\bibliography{ref}

\end{document}